\def\xmm{{\it XMM-Newton }}
\def\kev{{\it keV }}
\def\xspec{{\sc xspec }}
\def\chandra{{\it Chandra }}

\def\lledd{{${\rm L}/{\rm L}_{\rm Edd}$}}

\documentclass[12pt,preprint]{aastex}
\usepackage{graphicx}
\usepackage{mathptmx}
\usepackage{psfrag}
\usepackage{natbib}
 
\begin{document}
\title{\xmm observations of SDSS J143030.22$-$001115.1: an unusually
 flat spectrum AGN}

\author{S. Mathur \altaffilmark{1}, E.C. Golowacz \altaffilmark{1},
 R. Williams \altaffilmark{2}, R. Pogge \altaffilmark{1}, D. Fields
 \altaffilmark{3}, D. Grupe \altaffilmark{3} }


\altaffiltext{1} {Department of Astronomy, The Ohio State University,
 Columbus, OH 43210-1173 USA}
\altaffiltext{2} {Leiden Observatory, Leiden,
 Netherlands 2300RA }
\altaffiltext{3} {L.A. Pierce College, Woodland Hills, CA,
 91371 USA} 
\altaffiltext{4} {Pennsylvania State University, State College, PA, USA}

\begin{abstract} 
We present \xmm observations of the AGN SDSS 1430-0011. The low S/N
spectrum of this source obtained in a snap shot
\chandra observation showed an unusually flat continuum. With the follow
up \xmm observations we find that the source spectrum is complex; it
either has an ionized absorber or a partially covering absorber. The
underlying power-law is in the normal range observed for AGNs. The low
luminosity of the source during \chandra observations can be understood
in terms of variations in the absorber properties. The X-ray and optical
properties of this source are such that it cannot be securely classified
as either a narrow line Seyfert 1 or a broad line Seyfert 1 galaxy.
\end{abstract}
\maketitle

\section{Introduction}

Narrow--line Seyfert 1 galaxies (NLS1s) were initially classified by
Osterbrock \& Pogge~(1985) as a peculiar subset of active galactic
nuclei (AGNs) with strong, narrow H$\beta$ emission
(FWHM(H$\beta)\leq2000$ km s$^{-1}$), relatively weak [O\,III], and
strong optical Fe\,II emission. These spectral properties cause NLS1s to
stand out at one extreme end of the Boroson \& Green~(1992)
``Eigenvector 1'' (EV1). The physical driver behind EV1 is debated, but
is usually thought to correlate with the 
Eddington luminosity ratio, $L_{bol}/L_{Edd}$.  NLS1s then lie at the extreme
end of EV1 corresponding to high $L_{bol}/L_{Edd}$ (Pounds et al.~1995)
and also seem to have lower black hole masses than broad--line AGNs of
similar luminosities.

Subsequent X-ray studies found that many NLS1s have unusual X-ray
properties as well (e.g., Puchnarewicz et al.~1992). As a class, they
exhibit ultrasoft ($\Gamma
\mathrel{\hbox{\rlap{\hbox{\lower4pt\hbox{$\sim$}}}\hbox{$>$}}} 2.5$)
X-ray spectra compared to ``normal'' Seyfert 1s (Boller et al.~1996),
while some also show soft X-ray emission in excess of that expected from
a power law (Leighly~1999).  Kuraszkiewicz et al.~(2000) noted that this
ultrasoft X-ray emission may  be a consequence of high accretion
rates, and a correlation between $L_{bol}/L_{Edd}$ and $\Gamma$ is
indeed observed in the Grupe~(2004) sample.  Mathur~(2000) proposed that
the high accretion rate and low black hole mass indicate that NLS1s are
``young'' AGN; i.e. the central black holes are in an early stage of
their growth.  It was later found that NLS1s tend to fall below the
$M_{bh}-\sigma$ relation observed for broad--line Seyfert 1s supporting
this idea (Mathur et al. 2001) and bringing up the intriguing
possibility that AGN are ``born'' off of the $M_{bh}-\sigma$ relation and
eventually grow onto it through accretion (Grupe \& Mathur~2004; Mathur
\& Grupe 2005a,b; Watson et al. 2007).

Because of these properties, soft X-ray selection has 
proven to be an efficient technique for finding large numbers of NLS1s
(Grupe et al.~2004).  However, the aggregate X-ray properties of NLS1s
cannot be easily studied with soft X-ray--selected samples because these
necessarily exclude any NLS1s with harder X-ray emission, if they exist.
These issues were partially resolved
by the Sloan Digital Sky Survey (SDSS) with its homogeneous
selection criteria, and in particular the subsample of 
Williams et al.~(2002; hereafter W02) selected from the SDSS solely
on the basis of the Osterbrock \& Pogge~(1985) optical spectral criteria.
Indeed, the NLS1s from that sample which also appeared in the \emph{ROSAT}
All-Sky Survey (RASS) have on average flatter spectra than soft X-ray selected
samples (W02). 

A substantial number of NLS1s in the W02 sample should have been
detected in the RASS based on their optical brightness, but were not.
Short (2\,ks) observations of 17 of these X-ray--faint NLS1s were taken
with \emph{Chandra} in a follow--up study (Williams et al.~2004;
hereafter W04).  Some of these objects exhibit X-ray properties typical
of NLS1s, with $\Gamma\sim 2-3$ and X-ray to optical luminosity ratios
consistent with the RASS--detected NLS1s in W02.  However, four of the
NLS1s in the W04 sample are detected as unusually hard and faint
sources, with $\Gamma<2$ inferred from the \emph{Chandra} spectral fit
or hardness ratio (see below for details).  Additionally, those objects
with low $\Gamma$ tend to be much fainter in X-rays than the average for
RASS--detected NLS1s and the high--$\Gamma$ W04 NLS1s.

From works of W02 and W04 it is quite clear that NLS1s are much more
heterogeneous in their accretion properties than previously thought;
i.e.~even though NLS1s \emph{as a class} have high $L_{bol}/L_{Edd}$,
not all NLS1s do (see also Nikolajuk, Czerny, \& Gurynowicz 2009).  Some
NLS1s have steep X-ray spectra, but some do not.  Some NLS1s have strong
Fe\,II emission, but some do not.  It does appear that, for the most
part, NLS1s with large $L_{bol}/L_{Edd}$ have steep $\Gamma$
(Grupe~2004) and strong Fe\,II emission, and those are the objects whose
black holes are still growing (Mathur \& Grupe 2005a,b).  Of the four
flat--spectrum NLS1s in W04, two are very peculiar (SDSS
J143030.22$-$001115 and SDSS J1259$+$0102) with inferred $\Gamma=0.91$
and $0.25$ respectively, much too flat even for normal Seyfert 1
galaxies which have average $\Gamma=2$.

The original classification of SDSS J143030.22$-$001115.1 (SDSS
J1430-0011 here after) as a NLS1 was based on the SDSS spectrum
(W02). Bian, Cui \& Chao (2006) analyzed this spectrum again and found
that if you remove the narrow components of H$\beta$, the remaining
broad component has FWHM= 2600--2900 km s$^{-1}$ (dependent upon exact
modeling). Since the formal definition of NLS1 (Goodrich 1989) has a
maximum width of 2000 km s$^{-1}$, Bian et al. argue that SDSS 1430-0011
is not a NLS1. Even if SDSS J1430-0011 is a BLS1 or a NLS1 (discussed
further in \S 5.1), the \chandra spectrum with $\Gamma=0.91$ is still
peculiarly flat.

In principle, the low X-ray luminosities and low photon indices seen in
either of these AGNs could be caused by high intrinsic column density or
variability (though it is unlikely that both would be in such a low
state during both the RASS and \emph{Chandra} observations).
Unfortunately the individual \emph{Chandra} spectra of the four hardest
NLS1s contained too few counts to constrain both $N_H$ and $\Gamma$, but
a stacked spectrum of all four showed no evidence for strong absorption
($N_H < 2\times 10^{21} \rm{cm}^{-2}$ at the $2\sigma$ confidence
level).  A high S/N spectrum is clearly required to understand whether
the spectrum is truly flat or appears flat due to complexity.

We were awarded 25 ks of XMM time to obtain a high S/N spectrum of SDSS
 J1430-0011 (z=0.1032).  In \chandra observations, the source was found to be
 faint, with count rate CR=0.012.  W04 characterized its spectrum in
 therms of the hardness ratio, defined as $ HR\,=\,\frac{(H-S)}{(H+S)}$,
 where H and S are the net counts in the hard and soft bands,
 respectively (the hard band is defined as $2\,\kev<E<8\,\kev$ and the
 soft band as $0.4\,\kev<E<2\,\kev$).  They found $HR\,=\,-0.25$. They
 also fit a simple power-law to \chandra data and infer
 $\Gamma=0.92\pm0.64$. In the following, we present \xmm observations of
 this source

\section{Observations and Data Reduction}
The NLS1 SDSS J1430-0011 was observed with the \xmm European Photon
Imaging Camera (EPIC) pn and MOS detectors on 06 January 2008 for a
total of 25 ks.  All instruments were observed in extended full frame
mode with thin filters.

\subsection{Data Preparation}
The \textit{XMM-Newton} data were reduced using Science Analysis System
(SAS) 7.1.0. Light curves were initially produced for both EPIC pn and
MOS images in order to check for flaring high background. Periods of
high background are typically produced by protons in the Earth's
atmosphere with energies $\lesssim$ 100 \emph{keV} which are funneled
towards the detectors by the X-ray mirrors\footnote{\xmm Users' Handbook: http://xmm.esa.int/external/xmm\_user\_support/documentation/uhb/XMM\_UHB.html}.

Low background intervals were then produced by excluding data taken when
the full-field count rates exceeded 20 s$^{-1}$ in the pn and 2.5
s$^{-1}$ in the MOS.  Photon pile-up was also checked for, and was
determined not to be a problem.  Images were produced and binned into
square pixels of 4''.  Source and background data were then extracted.

For the EPIC pn data, the source counts were extracted with a circular
region of 20'' centered on the object.  Since the source was close to a
chip gap, background data was extracted from a source free circular
region of 20'' on the same CCD at about the same distance from the
readout node.  Furthermore, the source and background data were filtered
to include only single and double events (PATTERN 0-4).

For the MOS data, the source was extracted in a circular region of 20''
centered on the object.  The background was extracted in a surrounding
annulus of outer and inner radii 50'' and 25'' respectively.
Additionally, the source and background data were filtered to include
singles, doubles, triples, and quadruples  (PATTERN 0-12).

The SAS task \emph{backscale} was then run on both data sets to take
into account bad pixels and CCD boundaries.  A Redistribution Matrix
File and Ancillary Response File were then produced with the tasks
\emph{rmfgen} and \emph{arfgen} respectively. The final effective
exposure times and count rates for pn, MOS1 and MOS2 cameras are given
in Table 1.

\section{Data Analysis and Model Fitting}
The pn and MOS spectra were binned to have a minimum of 30 and 15 counts
per bin, respectively, using the FTOOLS program \textit{grppha} and then
analyzed using the \xspec 12.3.1 software package.  Joint fits were made
to the pn and MOS spectra.  We use solar abundances from Lodders (2003)
and photo-electric absorption cross-sections from Morrison \& McCammon
(1983).  Throughout the data analysis we use H$_0$=70 km s$^{-1}$
Mpc$^{-1}$, $\Omega_{\Lambda}=0.7$, $\Omega_m=(1-\Omega_{\Lambda})=0.3$.

We fit a variety of models, all of which are described in detail in
the subsections below.  In every model we included absorption by the
interstellar medium (ISM) of the Milky Way which was held fixed when
fitting the spectrum.  The Galactic column density toward SDSS
J1430-0011 is $N_H^{Gal}\,=\,3.15\times10^{20}\,cm^{-2}$ (Dicky and
Lockman 1990). The goodness of fit was determined through $\chi^2$
statistic. The errors quoted are for 90\% confidence for one interesting
parameter ($\Delta\chi^2=2.706$). In Table 2 we list only the acceptable
models with their fit parameters.


\subsection{Modeling the Data}
Before any modeling was done to the \xmm data, we applied the results
from the \chandra observation.  We fit our \xmm data with a simple power-law,
and fixed the photon index $\Gamma\,=\,0.92$ (derived in W04).  We
looked to see how the model behaved, and then made some assumptions
about the structure of the continuum and chose additional models
accordingly. The flat continuum derived from the \chandra data 
did not match the data at all and indicated that the true spectrum is
more complex than a simple power-law.

Outlined below are the models we used.  Joint fits were made to EPIC pn
and MOS spectra for $0.3\,\kev\,\leq E \leq\,10\,\kev$.  For simplicity,
we have  included their \xspec syntax.\\


{\bf Model 1: Simple Power-law}: This simple photon power-law model (at
 the redshift of the source) was fit with the \xspec command
 \textit{wabs(zpow)}.  The free parameters of the model were the photon
 index $\Gamma$, and the normalization. This was not a good fit, with
 $\chi^2=95.1$ for 60 degrees of freedom. Moreover, the best fit
 $\Gamma=2.3\pm 0.1$, is very different from the $\Gamma=0.92$ derived
 from the \chandra data. The true spectrum of the
 source, therefore, must be more complex than a simple power-law.  As
 shown in Figure 1, the fit left significant residuals at around 1
 keV. This is suggestive of an ionized absorber along the line of sight;
 such a model is discussed below.

{\bf Model 2: Intrinsic Absorber}: Apparent flatness of a continuum
can be caused by incorrect modeling of absorption. To investigate
whether this is the case, we next fitted the data with a model with
intrinsic absorption and a simple power-law continuum at the redshift of
the source (\xspec model \textit{wabs*zwabs(zpow)}). The free parameters
in this model were $N_H$, $\Gamma$, and the normalization. This model
did not result in a good fit either ($\chi^2=205$ for 59 degrees of
freedom). The residuals to the fit showed excess counts below about 1
keV. This implies that simple absorption by neutral matter at the source
is not the cause of the apparent flatness of the \chandra spectrum.

{\bf Model 3: Ionized Absorber}: This model consists of an ionized
 absorber with a simple power-law continuum at the redshift of the
 source (\xspec model \textit{wabs*absori(zpow)}). The free parameters
 of this model are $N_H$, $\Gamma$, the normalization, and absorber
 ionization state $\xi$\footnote{$\xi\,\equiv\,\frac{L}{N_eR^2}$, where
 L is the integrated luminosity from 5 eV to 300 keV, R is the radial
 distance from the source to the ionized material, and $N_e$ is the
 number density of electrons (Done, et al. 1992). The temperature was
 held fixed at $3\times10^4$~K, and the iron abundance was held fixed at
 the solar value.}.  This resulted in an acceptable fit (Table 2). The
 best fit vales of the parameters are $\Gamma=2.18^{+0.13}_{-0.14}$ and
 the column density of the ionized absorber $N_H= 4.3\times 10^{22}$
 cm$^{-2}$. The ionization parameter is $\xi=323^{+976}_{-212}$.  The
 intrinsic power-law slope is in the normal range observed for
 AGNs. Figure 2 shows the spectral fit of this model.

 {\bf Model 4: Partially Covering Absorber}: The apparently flat
 spectrum, the excess of counts at low energy in Model 2, and the dip in
 the residuals to the model 1 fit, are suggestive of a partially
 covering absorber, so we try this model next (\xspec model
 \textit{wabs*zpcfabs(zpow)}). The free parameters of this model are
 $\Gamma$, the normalization, $N_H$, and the covering fraction. The
 quality of this fit was similarly acceptable to that of Model 3 (Table
 2). The best fit vales of the parameters are $\Gamma=2.74\pm 0.20$ and
 $N_H= 4.47^{+4.78}_{-1.72}\times 10^{22}$ cm$^{-2}$. The covering
 fraction is $C_f=0.72^{+0.08}_{-0.13}$. The power-law slope is in the
 range observed for NLS1s. Figure 3 shows the spectral fit and the
 confidence contours are shown in figure 4.


{\bf Model 5: Disk Blackbody}: In an attempt to rule out other possible
scenarios, we looked at additional models that may also have been
representative of our source.

As discussed above, residuals to Model 2 show excess at low
energies. Similarly, if we fit the spectrum with a simple power-law for
E\,$\geq\,$2 \kev and extrapolate it down to lower energies, we again
see an excess in data. This upturn may indicate soft excess which is a
characteristic of NLS1s.  The cause of the soft excess in NLS1s is a
matter of debate (see e.g. Atlee \& Mathur 2009 and references
therein). For the purpose of this paper, however, we are only interested
in parametrizing the excess, so we looked at blackbody, comptonization,
and thermal plasma models described by Page, et al. (2004). Only the
disk blackbody model was deemed  acceptable based on fit
statistics, which is discussed here.

  A disk blackbody model describes the emission from an
 accretion disk as a series of blackbodies at different temperatures,
 which are emitting from different radii (see Mitsuda et al., 1984 and 
 Makishima et al., 1986.); \xspec model \textit{wabs(zpow+diskbb)}. The
 free parameters in this model were $\Gamma$, the
 normalizations\footnote{There are two normalizations.  One is the
 normalization of the power-law, which is the photon flux per unit
 energy at 1 keV.  The other normalization is associated with the disk
 blackbody, which is defined to be
 \Bigg($\frac{R_{in}/km}{D/10\,kpc}\Bigg)^2cos\theta$ where R is an
 ``apparent'' inner disk radius, D is the distance to the source, and
 $\theta$ is the angle of the disk ($\theta$\,=\,0 implies face on).},
 and the temperature at the inner disk radius. This model also fit the
 data well (Table 2). The resulting $\Gamma=1.73\pm0.16$ and the disk
 temperature is $0.118 \pm 0.02$ keV. 

A flat X-ray spectrum may also imply that the primary continuum is
suppressed and the spectrum has a reflection component. However, as
shown in figure 1, the characteristic ``hump'' of the reflection model
in the hard X-rays is not seen. SDSS 1430$-$0011 is also a type 1 AGN,
so unlikely to be completely dominated by a reflection
component. Therefore, we do not discuss this model in details. As
discussed below, it is clear that the intrinsic spectrum of the source
is not flat as inferred from the \chandra data. The \xmm data, however,
cannot distinguish among different complex models; fitting models that
are not physically motivated is, therefore, avoided. Moreover, the
utility of reflection models to infer underlying physical parameters is
limited by the unknown geometry of the reflector (Murphy \& Yaqoob
2009).

\section{Consistency with \chandra data}

As discussed above, the three models listed in Table 2 fit the \xmm data
well. The correct model of the spectrum of SDSS J1430-0011 should also
be consistent with the \chandra data.  In the disk black body model,
even the high energy power-law is as steep $\Gamma=1.73$, much steeper
than the $\Gamma=0.92$ derived from the \chandra data and there is the
black body excess at low energies. Thus, this model cannot lead to an
apparently flat spectrum during \chandra observation. Indeed the best
fit HR of this model is HR=$-0.61\pm 0.01$, much softer than
observed. For this reason, we do not discuss this model further.

To check the consistency with \chandra data for the other two acceptable
 models (models 3 \& 4) we calculate the predicted \chandra ACIS-S count
 rate over the 0.4-8.0 \kev range and the corresponding HR; we then
 compare these parameters with observations. The models were produced in
 \xspec and then exported to the Portable, Interactive Multi-Mission
 Simulator (PIMMS) software version 3.9\footnote{available at
 \texttt{http://heasarc.gsfc.nasa.gov/docs/software /tools/pimms.html}}
 for predicting count rates and HR. For the best fit parameters, the
 predicted count rates are CR=0.022 for model 2 and CR=0.0228 for model
 3. These are higher than the observed count rate. A change in column
 density between the \chandra and \xmm observations can lead to such a
 change in the count rate. Alternatively, the ionization parameter
 (model 3) or the covering fraction (model 4) could have been
 different. To investigate whether this is the case, we varied $N_H$,
 $C_f$, and $\xi$ in the subsequent analysis, while holding all other
 parameters of the models fixed.

In figure 5 we have plotted HR as a function of $N_H$ for the partial
covering model. The dotted lines are the predictions for a range of
$C_f$ as indicated. The thick solid regions on each curve correspond to
90\% confidence intervals for $N_H$. The dashed horizontal lines define
the 1$\sigma$ confidence interval of the \chandra HR. As can be clearly
seen from the figure, the best fit values of $N_H$ and $C_f$ are
inconsistent with the \chandra data. The best fit value of covering
fraction is inconsistent with \chandra data for any $N_H$. Covering
fraction between about 90\%--95\%, together with the best fit column density
are consistent with the \chandra data. However, the predicted count rate
for the $C_f=0.9$ model is CR=0.019, still above the observed value. On
the other hand, if $C_f=0.95$, then the predicted CR=0.012, as
observed. Thus, if a power-law with a partially covering absorber is the
correct description of the source spectrum, then the covering fraction
must have changed from 0.95 to 0.72 between \chandra and \xmm
observations.

In figure 6, we present a similar plot for the ionized absorber
model. The dotted curves are for a range of $\xi$ values, as
labeled. The dark solid lines encompasses the 90\% confidence range of
$N_H$ while the blue solid lines correspond to the parameter space
consistent with observed \chandra CR. Again, the best fit values of
$\xi$ and $N_H$ do not match the \chandra data. For the best fit
$\xi=323$, higher values of $N_H$, between $23.3<\log N_H<23.4$ match
the \chandra HR as well as CR. For the observed $N_H$, the ionization
parameter will have to be as low as $\xi=200$ to match the \chandra data.
Thus, if an ionized absorber model is the correct description of the
target spectrum, the column density had to be higher, or the ionization
parameter had to be lower, during the \chandra observation.

\section{Discussion}

The target of our \xmm observations, SDSS J141430-0011, showed an
unusually flat spectrum and low luminosity during \chandra observations.
The \xmm observations showed that the spectrum is complex; it can be
well described by an ionized absorber model or with a partially covering
absorber model. The best fit power-law slopes in both cases were steeper
($\Gamma= 2.18$ and $2.74$) respectively. Thus the intrinsic power-law
slope of the source is not flat. The apparent faintness and spectral
flatness during the \chandra observation can be explained if the
covering fraction during the \chandra observation was higher than that
found during the \xmm observation if the partial covering model is
correct. Alternatively, if the ionized absorber model is correct, then
either the absorber column density was higher or the ionization
parameter was lower during \chandra observations.

The X-ray luminosity of the source, however, is still low even after
proper modeling of the \xmm spectrum. With L(2--20 keV)$\approx
2.75\times 10^{42}$ erg s$^{-1}$, SDSS J1430-0011 falls below the Grupe
(2004) relation between X-ray power-law slope and luminosity. It's
Eddington luminosity ratio is also low \lledd $\approx 0.1$ for a black
hole mass of $\log M_{BH}=6.6$ calculated using H$\beta$ line width,
luminosity and the scaling relations and assuming that the bolometric
luminosity is about $9 \times$L(2--10 keV). It still falls below the W04
relation between optical and X-ray luminosity of RASS detected sample.

\subsection{Classification of SDSS 1430-0011}

Given all its properties, it's worth asking if SDSS 1430-0011 is a {\it
bona fide} NLS1 galaxy. As discussed in \S 1, Bian, Cui \& Chao (2006)
argue that it is not a NLS1. These authors arrived at this conclusion by
separating the narrow components of the H$\beta$ line from the broad
component and found that the width of the broad component is about 2800
km s$^{-1}$. They modeled the narrow component based on the [OIII]
lines. It was found that the [OIII] lines show blue asymmetry, a quality
more often found in NLS1s than in BLS1s (Mathur 2000; Komossa \& Xu
2007).  It is useful, therefore, to examine other properties of this
galaxy and compare them with the distributions found for NLS1s and
BLS1s. The S/N in the SDSS spectrum is low, but the high ionization
``coronal'' lines of Fe VII $\lambda 6087$ and Fe X $\lambda 6375$ are
possibly detected in the optical spectrum. NLS1s, with their steep X-ray
spectra often show strong coronal lines (e.g. Pfieffer et al. 200). The
observed X-ray power-law slope is $\Gamma$=2--2.3 (90\% confidence
range) for model 3 and 2.5--2.9 for model 4. This is consistent with the
range found for optically selected NLS1s of W04. It is also consistent
with the range of X-ray selected BLS1s in Grupe et al. (2004). The
FeII/H$\beta$ ratio of the source is $0.59\pm 0.17$. This lies in the
overlapping region between the peaks of NLS1s and BLS1s in the
distribution found by Grupe et al. (2004). At high X-ray luminosities,
NLS1s typically show steeper spectra than BLS1s. But at low
luminosities, as observed for our source, both NLS1s and BLS1s have
$\Gamma \approx 2.5$ (Grupe 2004; their figure 7).  Thus it appears that
SDSS 1430-0011 is at border line between NLS1s and BLS1s.

 SDSS 1430-0011, however, is not alone in this ``in-between''
 classification. IRAS 13349+2438 (Gallo 2006) also has
 FWHM(H$\beta$)$\approx 2800$ km s$^{-1}$, though its other properties
 are similar to NLS1s. WPVS 007 (Grupe, Leighly, \& Komossa, 2008)
 sometimes behaves like a NLS1, but sometimes it is undetected in
 X-rays.  Mrk 335 (Grupe et al. 2008) in its low state can be considered
 similar to SDSS 1430-0011. All these observations suggest that the region
 between NLS1s and BLS1s is murky. This in-between state can be
 temporary in some cases, but there are also sources which usually
 occupy this region. Sometimes, it manifests itself as a flat X-ray
 spectrum (e.g. SDSS 1430-0011), sometimes as complex hard X-ray spectrum
 (c.f. Gallo 2006), sometimes transient X-ray spectrum (e.g. WPVS 007), or
 simply with broad H$\beta$ (e.g. IRAS 13349).

\section{Conclusion}

SDSS 1430-0011 appeared to show an extremely flat spectrum in the
\chandra observation. Our subsequent \xmm observations show that its
intrinsic spectrum is steeper, with power-law slope in the normal
observed range. The spectrum, however, is complex, with either a
partially covering absorber or an ionized absorber, which must have
varied between the \chandra and \xmm observations. Based on its optical
and X-ray properties, it is hard to classify SDSS 1430-0011 as either a
NLS1 or a BLS1; it is at the border line between the two. There are
several AGNs in this ``in-between'' class; SDSS 1430-0011 is not unique.

This work is supported in part by the NASA grant NNX07AQ63G to SM. DG is
supported in part on NASA grant NNX07AH67G.

\newpage

\begin{deluxetable}{ccc}
\tabletypesize{\scriptsize}
\tablecaption{SDSS J1430-0011, Obs ID 05015402 details.}
\tablewidth{0pt}
\tablehead{
\colhead{Instrument} &
\colhead{ Exposure Time (ks)\tablenotemark{a}} &
\colhead{Count Rate} 
}
\startdata
 pn & 12.9 & $6.617\times10^{-2}\pm2.411\times10^{-3}$ \\
MOS 1& 15.6 & $1.44\times10^{-2}\pm1.019\times10^{-3}$ \\
MOS 2 & 15.5 & $1.647\times10^{-2}\pm1.078\times10^{-3}$ 
\enddata
\tablenotetext{a}{Effective  exposure times after data reduction. }\\
\end{deluxetable}

\begin{deluxetable}{lccccccc}
\tablewidth{0pt}
\tabletypesize{\scriptsize}
\tablecaption{Model Parameters\tablenotemark{a}}
\tablehead{
\colhead{Model Name\tablenotemark{b}} &
\colhead{$N_H$} &
\colhead{$\Gamma_1$} &
\colhead{Other parameter\tablenotemark{c}} &
\colhead{$\chi^2_{\nu}/\nu$} &
\colhead{$P_{\chi}(\chi^2;\nu)$} &
\colhead{$L_{2-10}$} &
\colhead{$F_{2-10}$} \\
\colhead{(\xspec syntax)} &
\colhead{($10^{22}$ cm$^{-2}$)} &
\colhead{} &
\colhead{(keV)} &
\colhead{} &
\colhead{} &
\colhead{($\times10^{42}$ erg s$^{-1}$)} &
\colhead{($\times10^{-14}$ erg s$^{-1}$ cm$^{-2}$)} \\
}
\startdata\\
 Model 3: Ionized absorber & $4.3_{-2.5}^{+9.8}$ & $2.18\pm0.14$ & $323^{+976}_{-212}$ &  $0.954/58$ & 0.575 & $2.43$ & $9.08$\\
 &&&&&&&  \\
Model 4: Partial covering & $4.47_{-1.72}^{+4.78}$ & $2.74_{-0.20}^{+0.21}$  & \ $0.72_{-0.13}^{+0.08}$ & $1.128/58$ & 0.235 & $2.12$ & $7.66$\\
 &&&&&&& \\ 
Model 5: Disk blackbody & \dots & $1.73\pm0.16$ & $0.118\pm0.02$ & $0.9236/58$ & 0.640 & $2.65$ & $10.1$\\
 & &&&&&& \\ 
\enddata
\tablenotetext{a}{This table only lists models that were considered as representative}
\tablenotetext{b}{Galactic absorption was included in all models and was held fixed.}
\tablenotetext{c}{Ionization parameter $\xi$ for Model 3, covering fraction $C_f$ for Model 4, disk temperature in keV for Model 5.} \\
\end{deluxetable}

\newpage

\begin{figure}
\includegraphics[width=8.5cm]{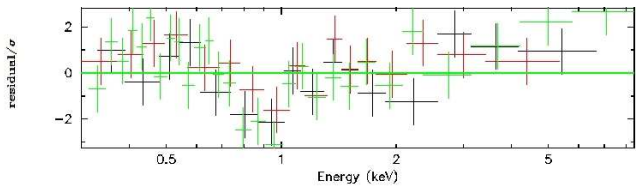}\\
\caption{Residuals ($\sigma$) to a fit with a simple power-law model (Model
 1). Notice the residuals at around 1 keV, suggestive of an ionized
 absorber. }
\end{figure}

\begin{figure}
\includegraphics[width=8.5cm]{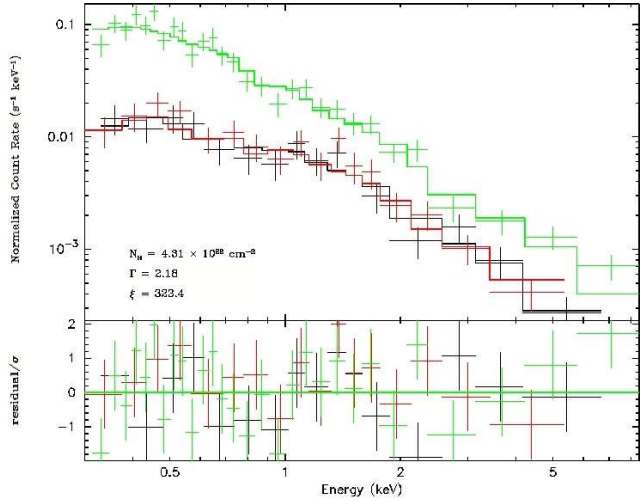}\\
\caption{A model of ionized absorber fits the \xmm data well (top panel). Residuals to the fit ($\sigma$) are plotted in the bottom panel. }
\end{figure}


\begin{figure}
\includegraphics[width=8.5cm]{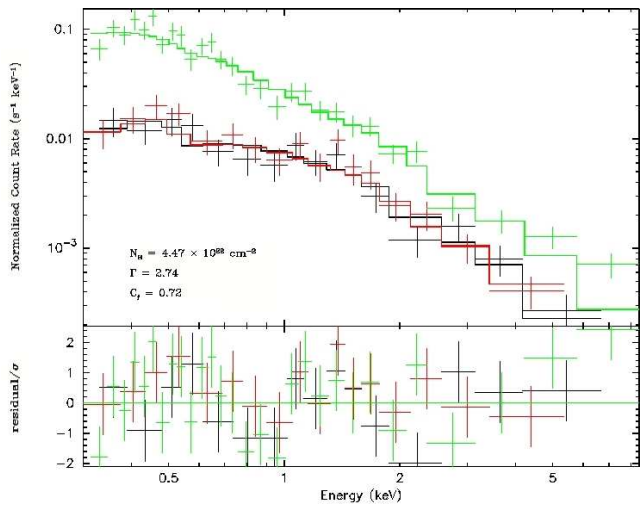}\\
\caption{Same as figure 2, but for a model with a partially covering absorber.}
\end{figure}

\begin{figure}
\includegraphics[width=8.5cm]{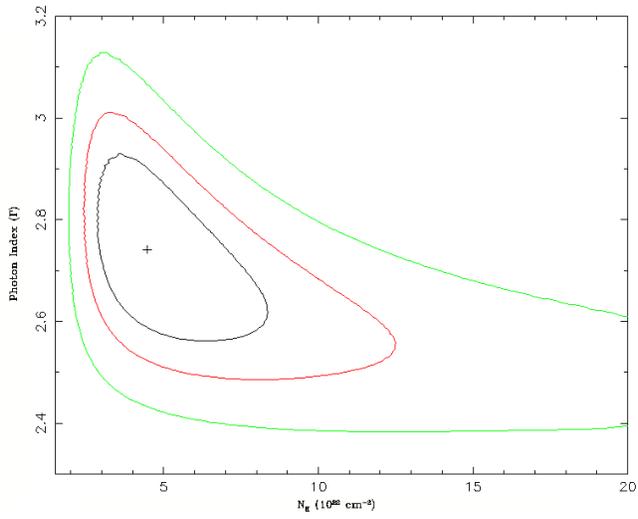}\\
\caption{Confidence contours of column density and power-law slope $\Gamma$ in
 Model 3. The best fit parameters and the one, two and 3$\sigma$
 intervals are shown.}
\end{figure}

\begin{figure}
\includegraphics[width=8.5cm]{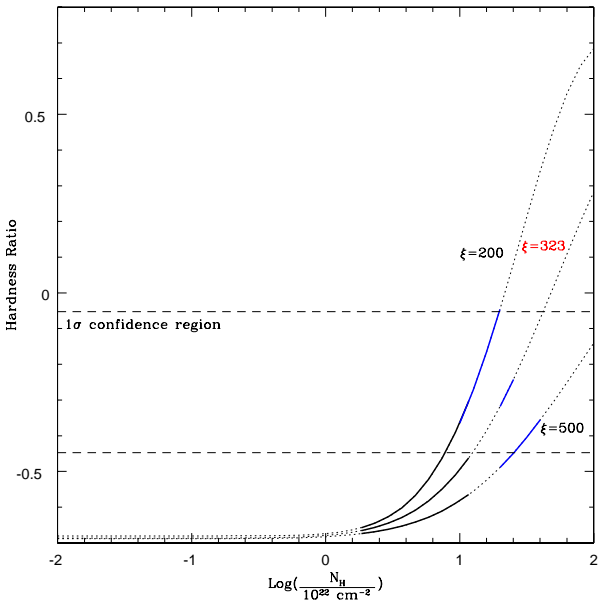}\\
\caption{Plot of hardness ratio vs. column density $N_H$ for the
 ionized absorber model. The dotted curves are for a range of $\xi$ as
 labeled. The middle curve is for the best fit $\xi$. The black solid
 lines are for the 90\% confidence range of observed $N_H$. The blue
 solid lines represent the \chandra count rate within 1$\sigma$
 confidence. The dashed horizontal lines are for the 1$\sigma$
 confidence range of \chandra hardness ratio. }
\end{figure}

\begin{figure}
\includegraphics[width=8.5cm]{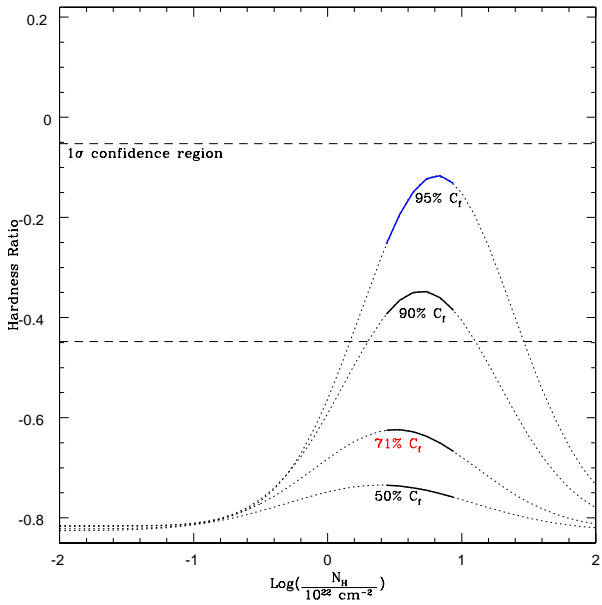}\\
\caption{Same as figure 6, but for the partial covering model. Here the
 dotted curves are for a range of covering fractions; the third curve
 from top is for the best fit C$_f$.  The solid blue lines are for the
 2--3$\sigma$ confidence interval of \chandra count rate. }
\end{figure}

\end{document}